\newcount\pageformat\pageformat=2  
\newcount\pssize

\ifnum\pageformat=2
 \documentclass[aps,prb,twocolumn,floatfix,showpacs,amsmath,amssymb]{revtex4}
\else
 \documentclass[aps,prb,preprint,floatfix,showpacs,amsmath,amssymb]{revtex4}
\fi
\usepackage{graphicx}

\begin{document}
\title{On a possibility of observation of quantum gravitational
effect by the cosmic laser ranging}

\author{B. F. Kostenko} \email{bkostenko@jinr.ru}

\affiliation{Joint Institute for Nuclear Research, Dubna\\
141980 Moscow region, Russia}

\date{\today }

\begin{abstract}
Physical interpretation of Lie algebra of de Sitter group is given
for Snyder theory of quantum space-time. Feasible means of its
experimental  validation are pointed out.
\end{abstract}

\maketitle

Up to date physical particles have not been introduced in Snyder
theory \cite{Snyder} in explicit form, in spite of many attempts to
apply the theory to solution of the  divergence problem in QED in
the sixties and modern using in Quantum Cosmology \cite{Han1, Han2}.
Strictly speaking, a generally accepted procedure of transformation
of de Sitter group generators into physical operators via
multiplication them by dimensional constants ($l_P=\sqrt {\hbar
G/c^3}$ or $\hbar$) is illegal because the generators contains space
points in mathematical momentum space, but not momentum coordinates
of physical particles. On the other hand, any physical measurement
is carried out over physical objects rather than space-time
coordinates itself, and discrete properties of space, suggested in
Snyder theory, should not be an exception.  Thus, symmetry group of
a space acquires physical meaning only after physical objects
(particles, fields etc.) are introduced. For example, the symmetry
of Minkowski space is reflected in symmetry of Lagrangian or
Hamiltonian containing coordinates of particles in space. In
addition,  we should take into account the matter-space coexistence
if we want our theory to be in line with General Relativity.

Here we follow an approach devised by E.~Wigner in a classical paper
\cite{Wigner}, where particles were defined as  irreducible
representations of Minkowski space group, only substituting for
place of particles' residence  de Sitter energy-momentum space. As
is well known, de Sitter group has two Casimir operators
\cite{Gursey}
$$
I_1=-\frac{1}{2 R^2} J_{ab}J_{ab}, \qquad I_2=- W_a W_a,
$$
where $J_{ab}$ are generators of  rotational group in
five-dimensional space  which cosmological de Sitter space-time is
embedded in, $R$ is radius of four-dimensional sphere corresponding
to this space-time, and
$$W_a=\frac{1}{8R} \varepsilon_{abcde}J_{bc} J_{de}. $$
In the limit $R \to \infty $, Casimir operators take on the form
$$
I_1 \to - P_{\lambda}P_{\lambda}=m^2, \qquad I_2 \to  m^2 s(s+1).
$$
Here $m$ and $s$ are mass and spin describing, according to Wigner,
a representation of Poincar\'e group in Minkowski space-time. Now,
replacing  space-time by energy-momentum with $R=\hbar/l_P$, we
obtain in the limit $R \to \infty $
$$
I_1 \to - X_{\lambda}X_{\lambda}=\chi^2, \qquad I_2 \to  \chi^2
s(s+1),
$$
where $X_{\lambda}$ describe {\it coordinates} of particles in
Minkowski space.

Thus, representations of de Sitter group in momentum space are
characterized not by $X_{\lambda}X_{\lambda}$ and $s$, but $I_1$ and
$I_2$ and states of particle should be designated as $\left| {I_1
I_2 } \right\rangle$. As far as $I_1$  does not correspond to a
certain space-time interval for $R=\hbar/l_P$, it is impossible to
attribute to the particle neither a certain distance from  the
origin of coordinates ($\chi^2<0$) nor a certain moment of time
($\chi^2>0$) in any reference frame. Furthermore, the very notion of
space-time interval allowing to distinguish cases $\chi^2<0$ and
$\chi^2>0$ is absent for $R=\hbar/l_P \ne \infty$. Therefore, a
rigorous distinction of rulers and clocks, which is characteristic
for special relativity, is lost in this case. Of course, this is an
imprint in physical states, $\left| {I_1 I_2 } \right\rangle$,  of
noncommutativity of space and time operators defined by Snyder in
empty de Sitter momentum space.

In fact, $\left| {I_1 I_2 } \right\rangle$ corresponds only to
reference vectors in state space of particles and may be used for
construction of more realistic particle states. To define these
vectors uniquely, we should choose a complete set of commutating
Hermitian operators. Actually, we can add to $I_1$ and $I_2$ only
one operator of 10 generators of de Sitter group since  they are
noncommuting. These operators are $\hat t$, $\hat x_i$, $\hbar {\hat
L}_i$, $\hbar {\hat M}_i$, $i=1,2,3$, where $\hat t$, $\hat x_i$ are
time and space operators of particle, $\hbar {\hat L}_i$ are
components of its angular momentum, and $\hbar {\hat M}_i$ are
generators of its boosts. If particle is in a state corresponding to
eigenvector of one of these generators, measurements of the others
will reveal stochastic results.

To describe such a stochasticity in detail we define the most
general expression for density operator,
$$
\hat \rho  = \sum\limits_{I,I'} {\left\langle {I} \right|} \hat \rho
\left| {I'} \right\rangle \left| {I } \right\rangle \left\langle {I'
} \right|,
$$
where $I$ is  multiindex, containing numerical values of $I_1$,
$I_2$ and an additional observable mention above. Now we may define
matrix elements of $\hat \rho$ on the surface of four-dimensional
hyperboloid,
$$\rho
(\xi ,\xi ') = \left\langle \xi  \right|\hat \rho \left| {\xi '}
\right\rangle  = \sum\limits_{I,I'} {\left\langle {I } \right|} \hat
\rho \left| {I'} \right\rangle \Psi _{I'}^* (\xi ')\Psi _{I} (\xi ),
$$
and operator of observable $\hat A$,
$$
\left\langle \xi  \right|\hat A\left| {\xi '} \right\rangle  =
\delta (\xi  - \xi ') A(\xi ).
$$
Then for mean value of $\hat A$ we obtain
$$
\left\langle {\hat A} \right\rangle  = {\rm Tr}(\hat \rho  \hat A) =
\sum\limits_{I,I'} {\left\langle I \right|} \hat \rho \left| {I'}
\right\rangle \int {\Psi _{I'} ^* (\xi )A(\xi )} \Psi _I (\xi )d\xi
.
$$

According to \cite{Kraus}, commutation relations
$$
\left[\hat x_i,\hat x_j\right] = i\frac{l_P^2}{\hbar}\hbar \hat L_k,
\qquad \left[ \hat t,\hat x_i\right] = i\frac{l_P^2}{\hbar c} \hbar
\hat M_i
$$
result in inequalities
$$
\Delta {x_i}\Delta {x_j} \ge \frac{1}{2}\frac{{{l_P^2}}}{\hbar
}\left| {\left\langle {\hbar {{\hat L}_k}} \right\rangle } \right|,
\qquad \Delta t\Delta {x_i} \ge \frac{1}{2}\frac{{{l_P^2}}}{\hbar
c}\left| {\left\langle {\hbar {{\hat M}_i}} \right\rangle } \right|
$$
in Snyder theory, and just these  relations may be  checked
experimentally.

We suggest that Snyder theory in a very general form described above
is applicable to any physical objects including cosmological ones,
associating $x_i$ with coordinates of their centers of mass. We may
expect that $\Psi _I (\xi )$ and $A(\xi )$ can be found
theoretically and that $s=0$ for $R \to \infty$. Density matrix
$\left\langle {I} \right| \hat \rho \left| {I'} \right\rangle $ is
in the general case a subject of difficult experimental and
theoretical investigations known as quantum state tomography.
Nevertheless, for some cosmological objects we can estimate value of
$\left| \left\langle \hat L_k \right\rangle \right|$ without
knowledge of quantum state $\hat \rho$. For example, using known
inertia moment of Earth, 8.042$\times$10$^{37}$ kg$\cdot$m$^2$, and
angular velocity of its daily rotation, we obtain $\left|
\left\langle \hat L_3 \right\rangle \right|=5.85\times 10^{33}$ kg
m$^2$s$^{-1}$. Then the uncertainty relation gives $\Delta x_1$,
$\Delta x_2 \approx$ 8.5 cm. Exact value of   moment inertia of Sun
is not known, and we can estimate its maximum value treating it as a
solid ball. Then $\Delta x_1$, $\Delta x_2 \le $ 1 km, so that
uncertainties of its c.m. position should be $ \sim$100 m in plane
transverse to axis of its axial rotation.

Modern cosmic geodesy exploits knowledge about coordinates of Earth
c.m. for definition of International Terrestrial Reference Frame
\cite{Altamimi}, in which it plays the role of the coordinates
origin (geocenter). The main idea of determination of the origin is
very simple: sets of coordinates of a satellite carry the
instantaneous  information about a geocenter position. In practice,
calculations require detailed data on terrestrial gravitational
potential, advanced troposphere modeling, corrections for antenna
thermal deformations and so on. However,  it is evident that
uncertainties of Earth coordinates predicted in Snyder theory should
be, this way or another, proportional to uncertainties of  its
satellite coordinates. Modern space geodetic techniques, such as
VLBI, SLR, GNSS and DORIS \cite{Altamimi}, allow to calculate these
coordinate with 1 cm precision. However, the calculations are
fulfilled on basis of smoothing and data averaging  over time and do
not contain information about instantaneous satellite coordinate.
For example, legacy theory \cite{Nesterov} and nowaday practice
require for gaining so-called normal points to fulfil averaging
several tens results of one-shot satellite laser ranging. Therefore,
for verification of the theory discussed here one should  make
precise measurements of Earth satellite coordinates using one-shot
laser ranging from special-purpose satellite (to reduce errors
caused by atmosphere effects).

The things stand similarly for the precise spacecraft observations
were done till now. Allan variance for Cassini, GALILEO and ULYSSES
shows that averaging was made over period $\tau = 1000$ s in this
case. Only separate measurements were made for GALILEO for $\tau =
1$ s. However, fluctuations less than 1 km could not be identified
even then.

In addition, quantum-gravitational effects, if they exist, can
manifest themselves in future cosmic experiments for gravity waves
detection. For example, eLISA  experiment does not provide fixation
of distances between spacecrafts in the constellation, in contrast
to DECIGO experiment, and this certainly will give rise to smearing
the interference pattern and to impossibility of gravity waves
detection.

\end{document}